\documentclass[conference]{IEEEtran}
\IEEEoverridecommandlockouts
\usepackage{cite}
\usepackage{amsmath,amssymb,amsfonts}
\usepackage{algorithmic}
\usepackage{graphicx}
\usepackage{textcomp}
\usepackage{xcolor}
\def\BibTeX{{\rm B\kern-.05em{\sc i\kern-.025em b}\kern-.08em
    T\kern-.1667em\lower.7ex\hbox{E}\kern-.125emX}}
\begin{document}

\title{Multi-Beam Symbol-Level Secure Communication for Hybrid Near- and Far-Field Communications
}

\setlength{\lineskiplimit}{0pt}
\setlength{\lineskip}{0pt}
\setlength{\abovedisplayskip}{6pt}   

\setlength{\abovedisplayshortskip}{6pt}
\setlength{\belowdisplayshortskip}{6pt}

\title{Multi-Beam Symbol-Level Secure Communication for Hybrid Near- and Far-Field Communications
}

\author{\IEEEauthorblockN{Wensheng Deng$^{\dagger}$, Bin Qiu$^{\dagger}$, and Wenchi Cheng$^{\dagger}$}\\[0.2cm]
\vspace{-10pt}

\IEEEauthorblockA{$^{\dagger}$State Key Laboratory of Integrated Services Networks, Xidian University, Xi'an, China\\
E-mail: \{\emph{23011210849@stu.xidian.edu.cn}, \emph{qiubin@xidian.edu.cn}, \emph{wccheng@xidian.edu.cn}\}}

\vspace{-20pt}
}

\maketitle

\begin{abstract}
This paper introduces a multi-beam secure communication scheme for mixed near-field and far-field (hNF) scenarios, primarily designed to address the challenges faced by sixth-generation (6G) networks in simultaneously managing near-field and far-field communications. This method significantly reduces the signal quality at eavesdroppers(Eves) while ensuring high-quality reception for legitimate users, effectively enhancing communication security. At the transmitter, this study employs multi-beam symbol-level directional modulation to ensure secure and reliable transmission in a mixed eavesdropping environment. Given the uncertainty about eavesdropper information, the transmission beamforming vectors are specially designed to meet specific symbol-level constraints, thereby ensuring effective reception for legitimate users. Experimental and simulation results demonstrate the effectiveness of our approach in improving secrecy performance, reducing transmission power, and enhancing energy efficiency, offering a practical solution to the security challenges faced by future wireless networks.
\end{abstract}

\begin{IEEEkeywords}
hybrid near- and far-field communications, physical layer security, symbol-level, multi-beam
\end{IEEEkeywords}

\section{Introduction}
To meet the escalating demand for higher data throughput in future wireless networks, research and implementation of ultra-massive multiple-input multiple-output (MIMO) systems, featuring hundreds of antennas, are progressing rapidly \cite{b1}\cite{b2}. This large-scale antenna use inevitably results in wireless communication operations within the near-field region. Unlike traditional far-field scenarios where the plane wave channel model is adequate, near-field communications are more accurately described by the spherical wave channel model, which incorporates both the direction and distance information of the receiver, allowing for focused array radiation patterns in free space (i.e., beam focusing) \cite{b3}. This capability enables near-field communications to utilize the new dimension of distance for enhanced signal precision, offering new opportunities in wireless communication.

Furthermore, the inherent broadcast nature of wireless channels makes the transmitted signals vulnerable to interception in hostile environments. Over the past decade, Physical Layer Security (PLS) has emerged as a critical complement to traditional cryptographic measures, widely recognized in the academic field \cite{b4}. The core principle of PLS involves leveraging the inherent randomness of wireless channels for secure transmissions. Directional Modulation (DM) stands out by using spatial degrees of freedom and antenna gain to focus transmissions on intended receivers while minimizing leakage to potential eavesdroppers \cite{b5}. The work by Lyu et al. introduces fractal orbital angular momentum (OAM) generation and detection schemes, providing new implementation avenues for directional modulation and further enhancing physical layer security \cite{b6}.

However, in practical scenarios, some legitimate users are located in the near-field region while others are in the far-field, creating complex challenges in interference management. Traditional techniques, such as beam focusing for near-field users (NU) and beamforming for far-field users (FU), must be adapted for integrated seamlessly into hybrid near- and far-field communication (hNFC) systems, as interactions between near and far fields can result in significant interference \cite{b7}.Given the inherent complexities of modern wireless networks, there is an urgent need to develop sophisticated interference management strategies that not only boost system performance but also ensure a consistent quality of service for all users \cite{b8}.The research by Cheng et al. discusses strategies for using adaptive finite blocklength coding in ultra-low latency wireless communications, crucial for improving service quality for both near-field and far-field users \cite{b9}.

To address these limitations of the previous works, this paper proposes a secure physical layer transmission within a hNF architecture. First, we utilize GPS technology to determine the positional and angular information of users, distinguishing between near-field and far-field users. Next, the transmitter employs a multi-beam directional modulation technique, designed to meet the symbol-level constraints of legitimate users, minimize the power of transmitted information, and ensure high-quality transmission for authorized recipients \cite{b10}. Finally, through simulations, we demonstrate the superiority of our approach in secure transmission within hNFC.
\section{SYSTEM MODEL}
\begin{figure}[htbp]
\centerline{\includegraphics[width=1\columnwidth]{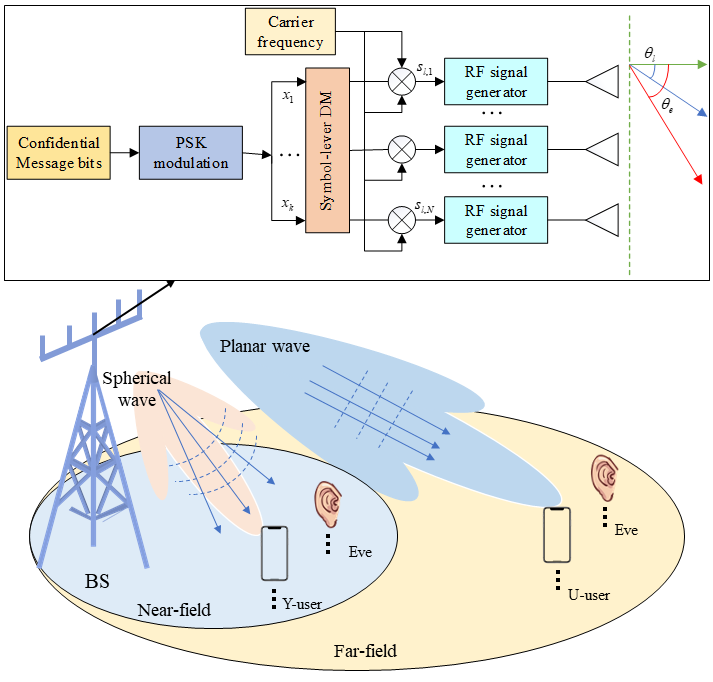}}
\caption{The multi-beam symbol-level communication.}
\label{fig1}
\end{figure}

As shown in Figure \ref{fig1}, we investigate a multi-beam symbol-level secure communication system within a hNF scenario. The system features a base station (BS) antenna serving $Y$ NUs and $U$ FUs. The BS is equipped with a transmitting array consisting of $N$ elements, each spaced $d_t$ apart. This BS transmits different information streams simultaneously to $K$ legitimate users, where $K = Y + U$, using a Uniform Linear Array (ULA). For generality, the first element of the transmitting array is designated as the reference element.

The positions of near-field users (NUs) and far-field users (FUs) relative to the base station (BS) are denoted by
$r_y$ and $r_u$, respectively, and are differentiated by the Rayleigh distance $Z$, where $Z = \frac{{2{D^2}}}{\lambda }$. Here, $D$ represents the array aperture and $\lambda$ the carrier wavelength. $r_y < Z$ and $r_u > Z$. the angles of departure  from the BS to NU $y$ and FU $u$ are given by $\phi_y$ and ${{\mathbf{\theta }}_u}$, respectively.
\subsection{Channel Model}
Firstly, we present the models for the far-field and near-field respectively.
\paragraph{Far-Field Channel Model} When the user is situated in the far field region of the BS (i.e., $r_u > Z$), the channel from  the BS to FU $u$ can be described by the following formula:
\begin{eqnarray}
{{\mathbf{h}}_{far}} = \sqrt N {h_{far}}{\mathbf{a}}({\theta _u}),
\label{eq1}
\end{eqnarray} 
where ${{\bf{h}}_{far}}$ denotes the complex-valued channel gain between the BS and FU $u$, and ${\mathbf{a}}({\theta _u})$ is the far-field steering vector, given by

\begin{equation}
\begin{split}
{\mathbf{a}}({\theta _u}) = \frac{1}{{\sqrt N }}{\left[ {1,{e^{ - j2\pi \frac{{{f_c}{d_t}\cos {\theta _u}}}{c}}}, \cdots ,{e^{ - j2\pi \frac{{{f_c}({N_t} - 1){d_t}\cos {\theta _u}}}{c}}}} \right]^H},
\end{split}
\label{eq2}
\end{equation}
where $c$ represents the speed of light, and $f_c$ denotes the carrier frequency.
\paragraph{Near-Field Channel Model} For the NU(i.e., $r_y < Z$), we adopt the spherical-wave propagation model.  Consequently, the near-field channel from the BS to NU m is characterized as follows:
\begin{eqnarray}
{{\bf{h}}_{near}} = \sqrt N {h_{near}}{\bf{b}}({\phi _m},{r_y}),
\label{eq3}
\end{eqnarray} 
where ${{\bf{h}}_{near}}$ is the complex-valued channel gain, ${\bf{b}}({\phi _y},{r_y})$ denotes the near-filed steering vector,given by
\begin{equation}
\begin{split}
{\bf{b}}({\phi _y},{r_y}) = \frac{1}{{\sqrt N }}{\left[ {{e^{ - j2\pi \frac{{{r_{y,0}} - {r_y}}}{\lambda }}}, \cdots ,{e^{ - j2\pi \frac{{{r_{y,N - 1}} - {r_y}}}{\lambda }}}} \right]^H},
\end{split}
\label{eq4}
\end{equation}
where ${r_{y}} = \sqrt {r_y^2 + d_t^2{n^2} - 2{r_y}{d_t}n\cos {\phi _y}}$ is the distance from the $n$-th antenna to NU $y$.

\subsection{Multi-beam symbol-level communication model}
In our communication system model, which exhibits hNF characteristics, we initially obtain the positional information of legitimate users through GPS technology. By determining the position $R$ of legitimate users relative to the BS, we compare it with the Rayleigh distance to differentiate between NUs and FUs.

Secondly, employing a multi-beam symbol-level precoding subset transmission method, as illustrated in Figure \ref{fig1}, The expression for the instantaneous baseband transmission signal ${{\mathbf{s}}_l}$ of a legitimate user is as follows \cite{b5}
\begin{eqnarray}
{{\mathbf{s}}_l} = \sum\limits_{k = 1}^K {{{\mathbf{w}}_k}{x_k}} ,
\label{eq5}
\end{eqnarray} 
where ${{\mathbf{w}}_k} \in {\mathbb{C}^{{N_t} \times 1}}$ is the beamforming vector controlling the transmission of the modulated symbol ${x_k}$ is an M-PSK modulated signal designed for legitimate user $k$, which satisfies $E\left[ {{{\left| {{x_k}} \right|}^2}} \right] = 1,\forall k \in {\rm K},{\rm K} \triangleq \left\{ {1,2, \cdots ,k} \right\}$.

Assuming that the legitimate users can achieve ideal frequency and phase synchronization, the down-converted signal received by legitimate user $k$, $\forall k \in K$ can be represented as:
\begin{eqnarray}
y({\theta _k}) = {{\mathbf{h}}^H}{{\mathbf{s}}_l}{{\mathbf{u}}_k} + {{\text{n}}_{l,k}},
\label{eq6}
\end{eqnarray} 
where ${\mathbf{h}}$ is the steering vector of the near- or far-field, ${{\mathbf{u}}_k} \in {\mathbb{C}^{N \times 1}}$, ${{\text{n}}_{l,k}}$ represents the additive white Gaussian noise (AWGN) with a zero mean and a variance of $\sigma _{l,k}^2$, and it is distributed as ${n_{l,k}} \sim N(0,\sigma _{l,k}^2)$.For simplicity, we will first discuss the far-field scenario; the operations for near-field and far-field are similar.

In the system described, it is assumed that there are multiple Eves, i.e., ${N_e} \geqslant 1$. Consequently, the orientation matrix of each Eve is expressed as:
\begin{eqnarray}
{\mathbf{H}}({\Theta _e}) \triangleq \left[ {{\mathbf{h}}({\theta _{e,1}}),{\mathbf{h}}({\theta _{e,2}}), \cdots ,{\mathbf{h}}({\theta _{e,{N_e}}})} \right].
\label{eq7}
\end{eqnarray} 

The Eve should use uncorrelated channels and adopt a distributed structure. So, the signal received by the Eve can be expressed as
\begin{eqnarray}
{\mathbf{y}}({\Theta _e}) = {{\mathbf{H}}^H}({\Theta _e}){{\mathbf{s}}_l} + {{\mathbf{n}}_e},
\label{eq8}
\end{eqnarray} 
where ${{\mathbf{n}}_e}$ is a complex AWGN and satisfies ${{\mathbf{n}}_e} \sim N(0,\sigma _e^2{{\mathbf{I}}_{{N_e}}})$.

Following the principles of information-theoretic PLS, secrecy performance is characterized by the secrecy capacity (SC). This capacity, denoted as $C_S$, is determined by the positive difference between the mutual information of the legitimate communication and the mutual information obtained through eavesdropping.
\begin{eqnarray}
{C_S} = {\left[ {{C_L} - {C_E}} \right]^ + },
\label{eqa3}
\end{eqnarray}  
where the mutual information between the BS and L/E is given by ${C_L} = \sum\limits_{k = 1}^K {\frac{1}{M}} {\log _2}\left( {1 + SN{R_{l,k}}} \right)$, ${C_E} = \sum\limits_{j = 1}^J {\frac{1}{M}} {\log _2}\left( {1 + SN{R_{e,j}}} \right)$.
\subsection{Secure transmission strategy}
To achieve secure wireless transmission, it is crucial to establish reliable communication between the transmitter and legitimate users while preventing information leakage. This paper proposes enhancing wireless security in a mixed eavesdropping environment by optimizing the beamforming vectors of the transmission array.

Focusing on relaxed phase conditions, we design the beamforming vector without requiring a fixed phase for the received signal. As long as the signal remains within the correct detection region, despite noise, the symbol information will be accurately recovered. This approach allows the phase of the received signal to vary within a relaxed region.

\begin{figure}[htbp]
\centerline{\includegraphics{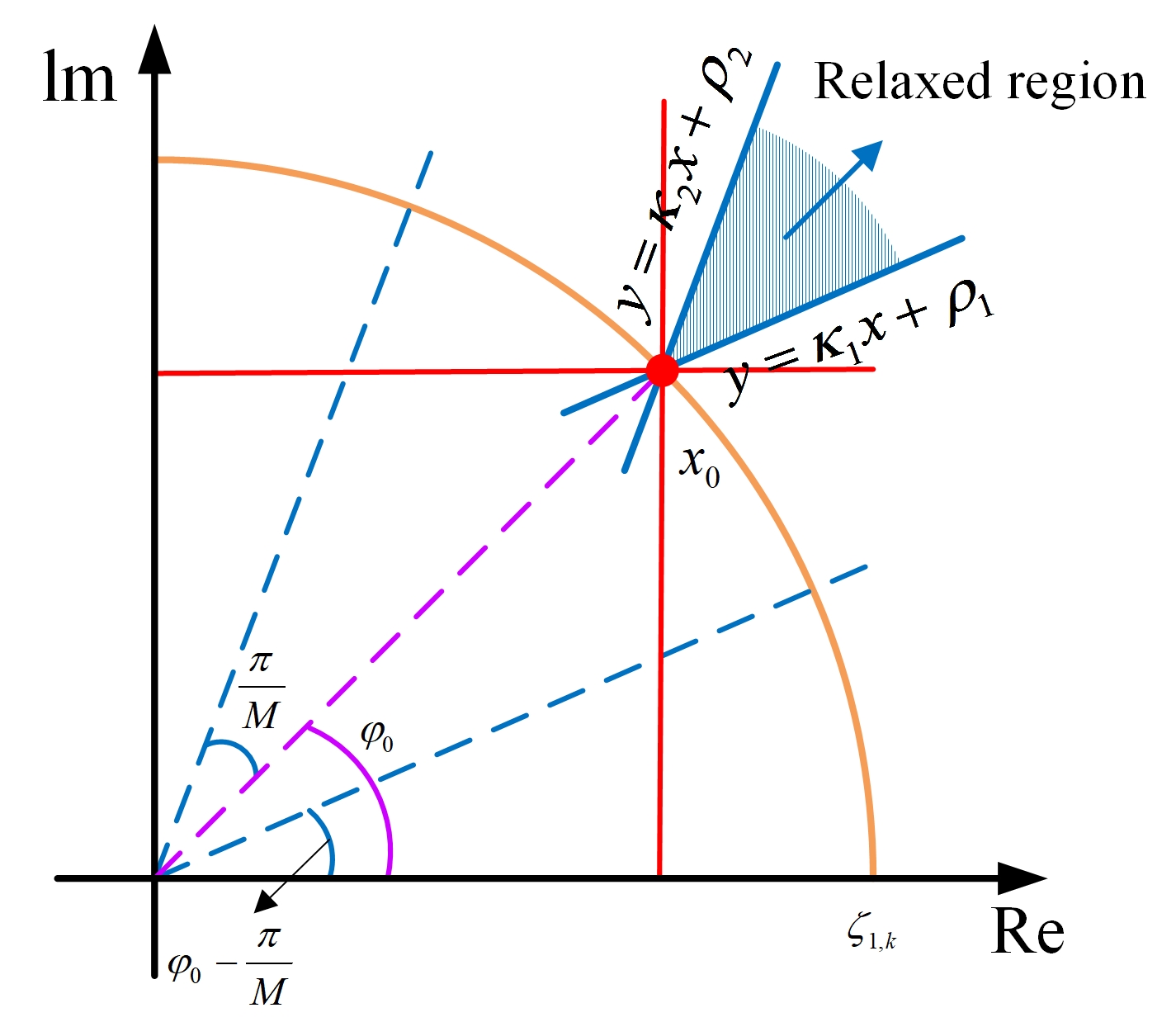}}
\caption{Structure of symbol level precoding with relaxed phase in the M-PSK modulation}
\label{fig2}
\end{figure}

Thus, we design the beamforming vector to allow the phase of the received symbol to vary within a predetermined phase range, where legitimate users can still correctly demodulate the symbol information. For M-PSK modulated symbols, with the phase of symbol ${x_0}$ as the reference phase ${\varphi_0}$, the relaxed region is defined as $\left[ {{\varphi_0} - \frac{\pi }{M}, {\varphi_0} + \frac{\pi }{M}} \right]$. This relaxed phase approach ensures that phases within this range are correctly demodulated, simplifying the design of the beamforming vector for secure transmission:
\begin{subequations}
\begin{flalign}
{\mathbf{P1}}:&\mathop {\min }\limits_{\left\{ {{{\mathbf{w}}_k}} \right\}_{k = 1}^K} \left\| {\sum\limits_k^K {{{\mathbf{w}}_k}{x_k}} } \right\|_2^2
\label{eq16a}\\
&{\mathbf{s}}{\mathbf{.t}}{\mathbf{.}}\left\| {{{\mathbf{h}}^H}\left( {{\theta _{l,k}}} \right)\left( {\sum\limits_k^K {{{\mathbf{w}}_k}{x_k}} } \right)} \right\|_2^2 \geqslant \zeta _k^2,\forall k \in K,
\label{eq16b}\\
& \arg \left\{ {{{\mathbf{h}}^H}\left( {{\theta _{l,k}}} \right)\left( {\sum\limits_k^K {{{\mathbf{w}}_k}{x_k}} } \right)} \right\} \geqslant {\varphi _0} - \frac{\pi }{M},\forall k \in K,
\label{eq16c} \\
&\arg \left\{ {{{\mathbf{h}}^H}\left( {{\theta _{l,k}}} \right)\left( {\sum\limits_k^K {{{\mathbf{w}}_k}{x_k}} } \right)} \right\} \leqslant {\varphi _0} + \frac{\pi }{M},\forall k \in K,
\label{eq16d}
\end{flalign} 
\end{subequations}
where ${\zeta _k}\in \mathbb{R}$ represents the specified SNR for the received signal of legitimate user $k$.
Constraint \eqref{eq16b} aims to protect the received signal of legitimate user $k$ by ensuring that the SNR of the received signal meets the specified requirements. Constraints \eqref{eq16c} and \eqref{eq16d} ensure that the phase of the received signal falls within the relaxed phase region. The relaxed region can be represented by the shaded sector in Figure \ref{fig2}, which is formed by two boundary parallel lines that intersect at the constellation point, spanning an angle of $\frac{{2\pi }}{M}$ in the direction opposite to the center, thus satisfying the linear constraints $[y = {k_1}x + {\rho _1},y = {k_2}x + {\rho _2}]$. 

To apply this to generic constellation points, consider the phase difference between ${\varphi _0}$ and ${\varphi _k}$. Similarly, define the overall beamforming vector ${\mathbf{v}} \triangleq \sum\limits_{k = 1}^K {{{\mathbf{w}}_k}{x_k}}$. Using basic geometric operations,and consider the constraints of all legitimate users optimization problem P1 is equivalent to the following optimization problem:
\begin{subequations}
\begin{flalign}
{\mathbf{P2}}:&\mathop {\min }\limits_{\mathbf{v}} \left\| {\mathbf{v}} \right\|_2^2
\label{eq17a}\\
&{\mathbf{s}}{\mathbf{.t}}{\mathbf{.}}{\mkern 1mu} {\text{Im}}\{ {\widetilde {\mathbf{H}}^H}({\Theta _l}){\mathbf{v}}\}  \geqslant {\kappa _1}{\text{Re}}\left\{ {{{\widetilde {\mathbf{H}}}^H}({\Theta _l}){\mathbf{v}}} \right\} + {{\mathbf{\rho }}_1},
\label{eq17b}\\
&\quad\,\, {\text{Im}}\{ {\widetilde {\mathbf{H}}^H}({\Theta _l}){\mathbf{v}}\}  \leqslant {\kappa _2}{\text{Re}}\left\{ {{{\widetilde {\mathbf{H}}}^H}({\Theta _l}){\mathbf{v}}} \right\} + {{\mathbf{\rho }}_2},
\label{eq17c} 
\end{flalign} 
\end{subequations}
where ${\kappa _1} = \tan ({\varphi _0} - \frac{\pi }{M})$, ${\kappa _2} = \tan ({\varphi _0} + \frac{\pi }{M})$, $\widetilde {\mathbf{H}}({\Theta _l}) = [\widetilde {\mathbf{h}}({{\mathbf{\theta }}_{l,1}}),\widetilde {\mathbf{h}}({{\mathbf{\theta }}_{l,2}}), \cdots ,\widetilde {\mathbf{h}}({{\mathbf{\theta }}_{l,K}})]$, $\widetilde {\mathbf{h}}({{\mathbf{\theta }}_{l,k}}) = {\mathbf{h}}({{\mathbf{\theta }}_{l,k}}) \circ {e^{j{\beta _k}}}$.Separate the real and imaginary parts of the complex number, and then proceed by
\begin{equation}
\begin{split}
\widetilde {\mathbf{H}}&({\Theta _l}){\mathbf{v}} = \operatorname{Re} \left\{ {{{\widetilde {\mathbf{H}}}^H}({\Theta _l})} \right\}\operatorname{Re} \{ {\mathbf{v}}\}  - \operatorname{Im} \left\{ {{{\widetilde {\mathbf{H}}}^H}({\Theta _l})} \right\}\operatorname{Im} \{ {\mathbf{u}}\} \\
& + j\left[ {\operatorname{Re} \left\{ {{{\widetilde {\mathbf{H}}}^H}({\Theta _l})} \right\}\operatorname{Im} \{ {\mathbf{v}}\}  - \operatorname{Im} \left\{ {{{\widetilde {\mathbf{H}}}^H}({\Theta _l})} \right\}\operatorname{Re} \{ {\mathbf{u}}\} } \right]
\end{split}
\end{equation}
This leads to
\begin{eqnarray}
\operatorname{Re} \left\{ {{{\widetilde {\mathbf{H}}}^H}({\Theta _l}){\mathbf{v}}} \right\} = \widetilde {\mathbf{H}}_1^T\widetilde {\mathbf{v}},
\label{eq13}
\end{eqnarray}
\begin{eqnarray}
\operatorname{Im} \left\{ {{{\widetilde {\mathbf{H}}}^H}({\Theta _l}){\mathbf{v}}} \right\} = \widetilde {\mathbf{H}}_2^T\widetilde {\mathbf{v}},
\label{eq14}
\end{eqnarray}
where $\widetilde {\mathbf{v}} = {\left[ {\operatorname{Re} \left\{ {{{\mathbf{v}}^T}} \right\},\operatorname{Im} \left\{ {{{\mathbf{v}}^T}} \right\}} \right]^T}$, $\widetilde {\mathbf{H}}_1^T = \left[ {\operatorname{Re} \left\{ {{{\widetilde {\mathbf{H}}}^H}({\Theta _l})} \right\}, - \operatorname{Im} \left\{ {{{\widetilde {\mathbf{H}}}^H}({\Theta _l})} \right\}} \right]$, $\widetilde {\mathbf{H}}_2^T = \left[ {\operatorname{Im} \left\{ {{{\widetilde {\mathbf{H}}}^H}({\Theta _l})} \right\},\operatorname{Re} \left\{ {{{\widetilde {\mathbf{H}}}^H}({\Theta _l})} \right\}} \right]$, By substituting equations \eqref{eq13} and \eqref{eq14} into optimization problem P2, we obtain
\begin{subequations}
\begin{flalign}
{\mathbf{P3}}:&\mathop {\min }\limits_{\widetilde {\mathbf{v}}} \left\| {\widetilde {\mathbf{v}}} \right\|_2^2
\label{eq18a}\\
&{\mathbf{s}}{\mathbf{.t}}{\mathbf{.}}\,{\mathbf{F}}\widetilde {\mathbf{v}} \geqslant {\mathbf{\rho }},
\label{eq18b}
\end{flalign} 
\end{subequations}
where
\begin{eqnarray}
{\mathbf{F}} = \left[ {\begin{array}{*{20}{c}}
  {\widetilde {\mathbf{H}}_2^T - {\kappa _1}\widetilde {\mathbf{H}}_1^T} \\ 
  {{\kappa _2}\widetilde {\mathbf{H}}_1^T - \widetilde {\mathbf{H}}_2^T} 
\end{array}} \right],{\mathbf{\rho }} = \left[ {\begin{array}{*{20}{c}}
  {{\rho _1}} \\ 
  { - {\rho _2}} 
\end{array}} \right].
\end{eqnarray}

The detailed iterative process for solving optimization problem P3 is outlined in Algorithm 1.
Below, we demonstrate that the iterative algorithm achieves stable convergence to the optimal solution.Assuming the initial values are set as ${{\mathbf{\xi }}^0}$ and ${\delta ^0}$, we use these values in Algorithm 1 to derive the optimal solution ${{\mathbf{\xi }}^ * }$ and ${\delta ^ * }$, which satisfies
\begin{eqnarray}
f({{\mathbf{\xi }}^ * },{\delta ^ * }) \leqslant f({{\mathbf{\xi }}^0},{\delta ^ * }) \leqslant f({{\mathbf{\xi }}^0},{\delta ^0}).
\label{eq15}
\end{eqnarray} 

Due to the objective function having a zero lower bound, each iteration in Algorithm 1 monotonically approaches the optimal value, thus ensuring that Algorithm 1 achieves stable convergence to the optimal solution.

It is worth noting that since the beamforming vector depends on the steering vector and the transmission symbols, it is necessary to update the beamforming vector at the symbol rate to ensure effective detection capabilities for legitimate users.
\begin{table}[t]
\begin{center}
\begin{tabular}{llr}
\hline
\textbf{Algorithm 1: }Iterative approach for the problem P3\\
\hline
Input: Pick up an initial point ${{\mathbf{\xi }}^0} \in {\mathbb{R}^{\left( {2N - K} \right) \times 1}}$, $\lambda  \in [0, + \infty )$,\\ and set $r = 1$.\\
1: Determine ${\delta ^{r - 1}}$ by substituting ${{\mathbf{\xi }}^{r - 1}}$ into\\ \quad ${\delta ^ * } = \max \{ {\mathbf{X}}\overline {\mathbf{H}} _1^T{\mathbf{B\xi }} - \zeta  \circ {{\mathbf{x}}_s},0\} $;\\
2: Determine ${{\mathbf{\xi }}^r}$ by substituting ${\delta ^{r - 1}}$ into \\${{\mathbf{\xi }}^ * } = {\left[ {\frac{{{{\left( {{\mathbf{TB}}} \right)}^T}{\mathbf{TB}}}}{\lambda } + {{\left( {{\mathbf{X}}\overline {\mathbf{H}} _1^T{\mathbf{B}}} \right)}^T}{\mathbf{X}}\overline {\mathbf{H}} _1^T{\mathbf{B}}} \right]^{ - 1}}{\left( {{\mathbf{X}}\overline {\mathbf{H}} _1^T{\mathbf{B}}} \right)^T}\left( {\zeta  \circ {{\mathbf{x}}_s} + \delta } \right)$\\
3: if $f({{\mathbf{\xi }}^{r - 1}}) - f({{\mathbf{\xi }}^r}) > \varepsilon $ then\\
4: \quad$r = r + 1$;\\
5: \quad Go to 1; \\
6: else\\
7: \quad Return ${\mathbf{\xi }}$;\\
8: end if
Output:Get the finally optional solution ${{\mathbf{\xi }}^ * }$.\\
\hline
\end{tabular}
\label{tab1}
\end{center}
\end{table}
\section{ Experimental Results and Analysis}
This section presents simulation results to verify the security performance of the proposed method. The simulation parameters are as follows: the transmission carrier frequency is set at 28 GHz. The system consists of a Uniform Linear Array (ULA) transmitter with N elements and K legitimate users. For simplicity, we assume there are four users, each located in both near-field and far-field regions. Assuming that all legitimate users have the same background thermal noise variance and specified SNR, unless otherwise specified, QPSK is selected as the baseband modulation signal. The signal attenuation factor is determined by the carrier frequency and transmission distance, following the free space electromagnetic wave propagation path loss model.
\begin{figure}[htbp]
\centerline{\includegraphics[width=0.75\columnwidth]{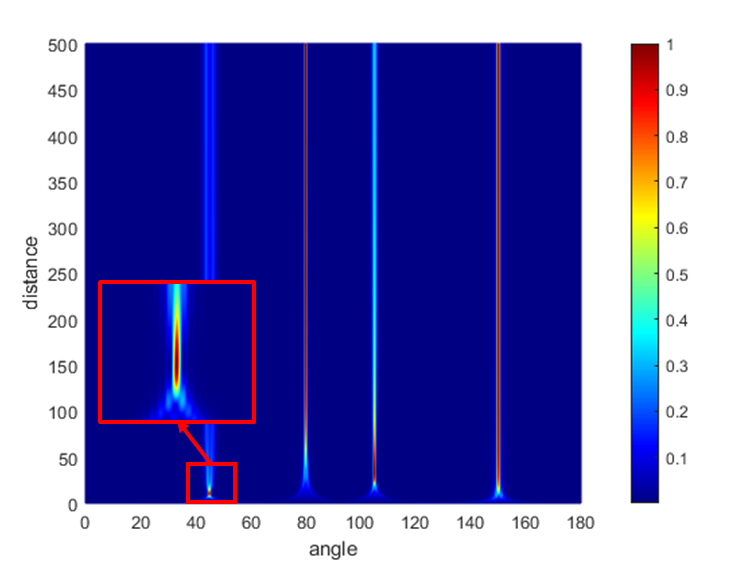}}
\caption{Normalized signal power spectrum}
\label{figa1}
\end{figure}

Firstly, Figure \ref{figa1} displays the normalized signal power spectrum, with $K = 4$, $N = 256$, ${d_{m,1}} = 10$, ${\phi _{m,1}} = 45$, ${d_{t,2}} = 390$, ${\theta _{t,2}} = 80$, ${d_{m,3}} = 35$, ${\phi _{m,3}} = 105$, ${d_{t,4}} = 420$, and${\theta _{t,4}} = 150$. it illustrates the received spatial distribution of energy, and as expected, the radiation patterns for the four users indicate the required directions, with near-field users also pinpointing the desired positions. The beamforming vectors we designed meet the requirements. Regarding the power requirements for LU reception, NF ensures physical layer security in terms of both angle and distance dimensions, achieving 'point' security transformation. FU ensures security in the angle dimension. In summary, the proposed method effectively suppresses interference signals and achieves reliable transmission between the transmitter and legitimate users according to specified communication quality, while reducing the possibility of Eves intercepting confidential information, thus achieving secure transmission in a hybrid environment.

\begin{figure}[htbp]
\centerline{\includegraphics[width=0.58\columnwidth]{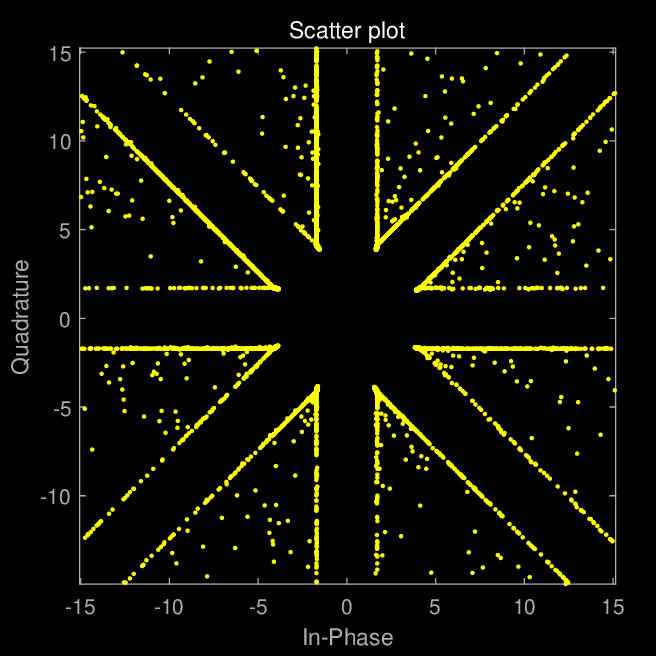}}
\caption{The 8-PSK received symbol constellation diagram}
\label{figa2}
\end{figure}

Figure \ref{figa2} depicts the 8-PSK modulated symbol constellation diagram received by legitimate users. As illustrated, the relative geometric shape of the noise-free received symbol constellation complies with the constraints of relaxed phase. The proposed design method consumes less transmission power while ensuring that all legitimate users receive signals at the specified SNR. This is achieved by transforming inter-symbol interference into beneficial power that aids other symbols. Moreover, as the proposed method requires less transmission power to ensure reliable reception by legitimate users, reducing the transmission power under a fixed total output enhances confidentiality.

\begin{figure}[htbp]
\centerline{\includegraphics[width=0.8\columnwidth]{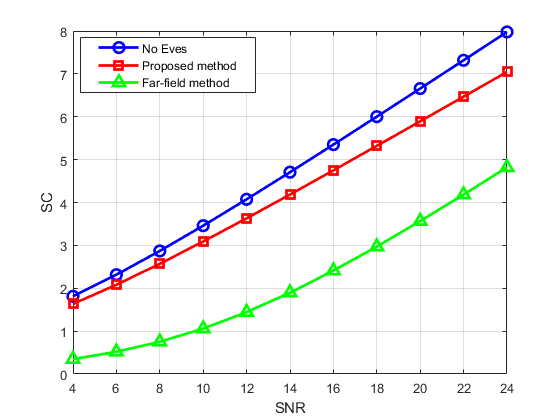}}
\caption{Secrecy Capacity as a Function of SNR}
\label{figa3}
\end{figure}

Figure \ref{figa3} illustrates the relationship between the SNR and SC. The graph depicts three lines, each representing the secrecy capacity under different conditions. The blue line shows the secrecy capacity in the absence of Eves, increasing with the SNR and representing the maximum achievable secrecy capacity in an ideal, threat-free environment. The red line corresponds to the method proposed in this study, which also demonstrates an increase in secrecy capacity with rising SNR, and the gap between this and the far-field method widens as the SNR increases. This proposed method surpasses the far-field method at all SNR points but falls below the ideal no-eavesdropper scenario, indicating that while it effectively enhances secrecy capacity, it remains susceptible to eavesdropping activities. The green line represents the far-field method, which exhibits significantly lower secrecy capacity at lower SNRs compared to the other two scenarios but steadily increases as SNR improves. However, across the entire range of SNR, the performance of the far-field method is inferior to that of the method presented in this paper.
\section{ Conclusion}
This paper addresses the issue of secure wireless transmission in mixed near-field and far-field environments. Specifically, by determining the target's position and angular information via GPS, the proposed method utilizes a multi-beam symbol-level directional modulation transmission subset. It optimizes the beamforming vectors to minimize transmission power while satisfying symbol-level constraints for each legitimate user. The method is efficiently supported by closed-form solutions, facilitating subsequent engineering implementations. Simulation results demonstrate that this method is an effective approach for achieving secure wireless transmission in mixed near-field and far-field environments.

\vspace{12pt}
\color{red}

\end{document}